\documentclass[showpacs,preprintnumbers,amsmath,amssymb]{revtex4}
\begin{document}

\title{Gravitational Corrections to the Energy-Levels of a Hydrogen Atom\footnote{This work was supported  by National Natural Science Foundation of
China  10435080 and 10575123, Chinese Academy of Sciences
Knowledge Innovation Project under Grant No. KJCX-SYW-N2 and
KJCX2-SW-N16.}}

\author{Zhao Zhen-Hua $^{1,}$\footnote{Corresponding author:  zhaozhenhua@impcas.ac.cn},
  Liu Yu-Xiao$^{2}$, Li Xi-Guo $^{3}$ }

\affiliation{ $^{1}$Institute of Modern Physics,Chinese Academy of
Sciences,
 Lanzhou 730000, China\\
$^{2}$Institute of Theoretical Physics, Lanzhou University,
 Lanzhou 730000, China \\
$^{3}$Institute of Modern Physics,Chinese Academy of Sciences,
 Lanzhou 730000, China\\
}


\begin{abstract}
The first order perturbations of the energy levels of a hydrogen
atom in central internal gravitational field are investigated. The
internal gravitational field is produced by the mass of the atomic
nucleus. The energy shifts are calculated for the relativistic
$1S$, $2S$, $2P$, $3S$, $3P$, $3D$, $4S$ and $4P$ levels with
Schwarzschild metric. The calculated results show that the
gravitational corrections are sensitive to the total angular
momentum quantum number.

\end{abstract}


\maketitle

\section{Introduction}

The study of gravitational fields interacting with spinor fields
constitutes an important element in constructing a theory that
combines quantum physics and gravity. For this reason, the
investigation of the behavior of relativistic particles in this
context is of considerable interest. It has been known that the
energy levels of an atom placed in an external gravitational field
will be shifted as a result of the interaction of the atom with
space-time curvature see
Refs.~\cite{Parker,Parker2,Parker3,Fischbach} for examples. And
the geometric and topological effects  lead to shifts in the
energy levels of a hydrogen atom are considered in
Ref.~\cite{Marques}.

Recently, there has been a dramatic increase in the accuracy of
experiments that measure the transition frequencies in hydrogen.
The most accurately measured transition is the $1S-2S$ frequency
in hydrogen; it has been measured with a relative uncertainty of
25 Hz ($\triangle f / f_0=1.0\times10^{-14}, f_0
=2466\;\text{THz}$) \cite{Fischer,Kolachevsky}, an order of
magnitude larger than the natural linewidth of 1.3 Hz natural
width of the $2S$ level \cite{Cesar,Killian}. Indeed, it is likely
that transitions in hydrogen will eventually be measured with an
uncertainty below 1 Hz \cite{Jentschura}. Though that accuracy can
not  explore the gravitational effect produced by the hydrogen
atom nucleus, with the progress of experiments we can detect the
gravitational  effect.

In this paper we investigate another previously neglected
gravitational effect of the energy-level shifts of a hydrogen atom.
This is to give some explicit values for energy-level shifts of a
hydrogen atom by the general relativistic effect with
Schwarzschild metric. And the difference with
Refs. \cite{Parker,Parker2,Parker3,Fischbach,Marques} is that the
gravitational field in this paper is not a external field but
produced by the mass of hydrogen atom nucleus. To our knowledge no
one has given explicit values for energy-level shifts of a hydrogen
atom with gravitational corrections. Although the effect is very
small, but it also has the physical significance as a test of
general relativity at the quantum level.

This paper is organized as follows: In Sec. \ref{GENERALLY
COVARIANT} we review the formalism  of the generally covariant
Dirac equation in curved space-time.  In Sec. \ref{SPINOR
CONNECTIONS}  we give the tetrad and spinor connections with
Schwarzschild metric. The gravitational perturbation of
relativistic $1S$ level is calculated in Sec. \ref{GRAVITATIONAL
PERTURBATION}.  The summary and discussion are given in Sec.
\ref{SUMMARY}.

\section{Generally Covariant  Dirac Equation in Curved
Space-Time}\label{GENERALLY COVARIANT}

To write the generally covariant Dirac equation in curved
space-time with metric $g_{\mu\nu}$, one first introduces the
spinor affine connections
$\omega_\mu=\frac{1}{2}\omega_\mu^{ab}I_{ab}$, where $I_{ab}$ are
the generators of $SO(4)$ group, whose spinor representation is
\begin{eqnarray}
  I_{ab}= \frac{1}{4}(\gamma_a \gamma_b -\gamma_b \gamma_a).
\end{eqnarray}
Here $\gamma_a$ are the Dirac-Pauli matrices with the following
relation
\begin{eqnarray}
  \gamma_a \gamma_b +\gamma_b
  \gamma_a=2\eta_{ab}\label{7},
\end{eqnarray}
and
\begin{eqnarray}
  \gamma_0^\dag&=&-\gamma_0,\quad \gamma_i^\dag=\gamma_i \qquad
  (i=1,2,3),\\
  \gamma_0&=&i\beta, \quad \quad  \gamma_i=-i
  \beta\alpha_i.\\
  \alpha_{i}&=& \left(
  \begin{array}{cc}
   0&\quad\sigma_i\\
   \sigma_i&\quad 0
   \end{array} \right),\quad
\beta= \left(
  \begin{array}{cc}
   I&\quad 0\\
   0&\quad -I
\end{array} \right),
\end{eqnarray}
where $I$ is the  $2\times2$ identity  matrix,
$\eta^{ab}=\eta_{ab}=\text{diag}(-1,1,1,1)$ is the Minkowski
metric tensor, and  $\sigma_i$ are the standard Pauli matrices
\begin{eqnarray}
  \sigma_1= \left(
  \begin{array}{cc}
   0&\quad 1\\
   1&\quad 0
\end{array} \right),\quad
\sigma_2= \left(
  \begin{array}{cc}
   0&\quad -i\\
   i&\quad 0
\end{array} \right),\quad
\sigma_3= \left(
  \begin{array}{cc}
   1&\quad 0\\
   0&\quad -1
\end{array} \right).
\end{eqnarray}
$\omega_\mu^{ab}$ is defined by the vanish of the generalized
covariant derivative \cite{Kibble,Loll} of the tetrad (or
vierbein) field \cite{Poisson} $e^{(a)}_{\quad\mu}(x)$
\begin{eqnarray}
  D_\mu e^{(a)}_{\quad\nu}
  &=&\partial_\mu e^{(a)}_{\quad\nu}-\Gamma^{\lambda}_{\mu\nu}e^{(a)}_{\quad\lambda}-\eta_{bc}\omega_{\mu}^{ab}e_{\quad\nu}^{(c)}\nonumber\\
  &=&\nabla_\mu e^{(a)}_{\quad\nu}-\eta_{bc}\omega_{\mu}^{ab}e_{\quad\nu}^{(c)}
  \equiv0 , \label{Omegamu}
\end{eqnarray}
where the  tetrad field $e^{(a)}_{\quad\mu}(x)$ and it's inverse
$e_{(a)}^{\quad\mu}(x)$ satisfy the following equations
\begin{eqnarray}
  &&g_{\mu\nu}(x)=\eta_{ab}e^{(a)}_{\quad\mu}(x)
  e^{(b)}_{\quad\nu}(x)\label{1},\\
  &&e^{(a)}_{\quad\mu}(x)e_{(b)}^{\quad\mu}(x)=\delta^{a}_{b}, \qquad (\mu, \nu, a,b=0, 1, 2,
  3)
\end{eqnarray}
$\mu,\nu$ are the space-time indices lowered with the metric
$g_{\mu\nu}$, and $a,b$ are the Lorentz group indices lowered with
$\eta_{ab}$.

One also needs to introduce generalized Dirac-Pauli matrices
$\Gamma_\mu(x)=e^{(a)}_{\quad\mu}(x)\gamma_{a}$, which satisfy the
equation \cite{Parker2}
\begin{eqnarray}
 \Gamma_\mu(x)\Gamma_\nu(x)+\Gamma_\nu(x)\Gamma_\mu(x)=2
 g_{\mu\nu}(x).
\end{eqnarray}
The covariant derivative acting on a spinor field $\psi$ is then
\begin{eqnarray}
  D_\mu \psi=\partial_\mu \psi -\omega_\mu \psi,\label{covariant derivative}
\end{eqnarray}
and  the generally covariant form of the Dirac
equation\cite{Fischbach} in pure gravitational field is
\begin{eqnarray}
  \Gamma^\mu(x) D_\mu \psi(x) + \frac{mc}{ \hbar} \psi(x)=0, \label{DiracEq1}
\end{eqnarray}
where $\Gamma^\mu(x)=g^{\mu\nu}\Gamma_\nu(x)$, $m$ is the mass of
spinor particles.

For an electron near the atomic nucleus one needs to consider the
effect of the electromagnetic vector potential $A_\mu$, here
$A_\mu$ satisfy the Maxwell equations \cite{Parker2,Misner}
\begin{eqnarray}
g^{\lambda\sigma}\nabla_{\lambda}\nabla_{\sigma}
A_{\mu}-R_{\mu}^{\;\;\nu}A_{\nu}=-4\pi J_{\mu},
\end{eqnarray}
where $J_{\mu}$ is the current vector. So the covariant derivative
acting on a spinor field should be rewritten as
\begin{eqnarray}
  D_\mu \psi&=&(\partial_\mu -\omega_\mu -i q A_\mu)\psi.
\end{eqnarray}
Then the generally covariant form of the Dirac equation in
gravitational and electromagnetic fields is
\begin{eqnarray}
  \Gamma^\mu(\partial_\mu -\omega_\mu -i q A_\mu)\psi(x) +\frac{mc}{ \hbar}
  \psi(x)=0.\label{DiracEq2}
\end{eqnarray}

\section{Spinor Connections in The Schwarzschild
Space-Time}\label{SPINOR CONNECTIONS}

In what follows, we will calculate the spinor connections in a
Schwarzschild spacetime. The line element corresponding to the
spacetime is given by
\begin{eqnarray}
  ds^2&=&-g_{\mu\nu}dx^{\mu}dx^{\nu} \nonumber \\
      &=& c^{2}\left(1-\frac{R_s}{r}\right)dt^2
          -\frac{1}{1-\frac{R_s}{r}}dr^2
          -r^{2}d\theta^2
          -r^2 \sin^{2}\theta d\phi^2,\label{metric}
\end{eqnarray}
where $R_s=2GM/r$. With the time gauge
conditions \cite{Schwinger,Maluf} $e^{(0)}_{\quad i}=0$ and
$e_{(i)}^{\quad 0}=0$, the tetrad field $e^{(a)}_{\quad\mu}$ is
given as follows:
\begin{eqnarray}
e^{(a)}_{\quad\mu}=\left(
\begin{array}{cccc}
 \sqrt{1-\frac{R_s}{r}} & 0 & 0 & 0 \\
 0 & \frac{\sin\theta\cos\phi}{\sqrt{1-\frac{R_s}{r}}} & r\cos\theta\cos\phi & -r\sin\theta\sin\phi \\
 0 & \frac{\sin\theta\sin\phi}{\sqrt{1-\frac{R_s}{r}}} & r\cos\theta\sin\phi & r\sin\theta\cos\phi \\
 0 & \frac{\cos\theta}{\sqrt{1-\frac{R_s}{r}}} & -r\sin\theta & 0
\end{array}
\right).
\end{eqnarray}
Taking the approximation $\sqrt{1-\frac{R_s}{r}} \cong
1-\frac{R_s}{2 r}$, we have
\begin{eqnarray}
e^{(a)}_{\quad\mu}=\left(
\begin{array}{cccc}
 1-\frac{R_s}{2 r} & 0 & 0 & 0 \\
 0 & \frac{\sin\theta\cos\phi}{1-\frac{R_s}{2 r}} & r\cos\theta\cos\phi & -r\sin\theta\sin\phi \\
 0 & \frac{\sin\theta\sin\phi}{1-\frac{R_s}{2 r}} & r\cos\theta\sin\phi & r\sin\theta\cos\phi \\
 0 & \frac{\cos\theta}{1-\frac{R_s}{2 r}} & -r\sin\theta & 0
\end{array}
\right).\label{tetrad}
\end{eqnarray}
From  Eq. (\ref{Omegamu}), it follows
\begin{eqnarray}
  \omega^{ab}_\mu=(\nabla_\mu e^{(a)}_{\quad\nu})e^{(b)}_{\quad\lambda}
  g^{\lambda\nu} ,\label{affine}
\end{eqnarray}
and
\begin{eqnarray}
  \omega_\mu = \frac{1}{2}\omega^{ab}_\mu I_{ab}
             = \frac{1}{2} I_{ab}(\nabla_\mu e^{(a)}_{\quad\nu})
                           e^{(b)}_{\quad\lambda} g^{\lambda\nu}
            \approx \frac{1}{2} I_{ab}(-\Gamma^{\rho}_{\mu\nu}) e^{(a)}_{\quad\rho}
                           e^{(b)}_{\quad\lambda} g^{\lambda\nu}.
            \label{omega}
\end{eqnarray}
 Thus using Eqs. (\ref{metric}), (\ref{tetrad}), (\ref{affine})and (\ref{omega}), we obtain the
 explicit  expressions of the nonzero components of spinor connections
 \begin{eqnarray}
 \omega _0=
 \left(
\begin{array}{cccc}
 0 & 0 & -\frac{R_s \cos\theta}{4 r^2} & -\frac{R_s \sin\theta\; e^{-i \phi}}{4 r^2} \\
 0 & 0 & -\frac{R_s \sin\theta\; e^{i \phi}}{4 r^2} & \frac{R_s \cos\theta}{4 r^2} \\
 -\frac{R_s \cos\theta}{4 r^2} & -\frac{R_s \sin\theta e^{-i \phi}}{4 r^2} & 0 & 0 \\
 -\frac{R_s \sin\theta\; e^{i \phi}}{4 r^2} & \frac{R_s \cos\theta}{4 r^2} & 0 & 0
\end{array}
\right) ,
\end{eqnarray}
\begin{eqnarray}
\omega _2=\left(
\begin{array}{cccc}
 0 & \frac{(r-R_s) e^{-i\phi}}{2 r-R_s} & 0 & 0 \\
 -\frac{(r-R_s) e^{i\phi}}{2 r-R_s} & 0 & 0 & 0 \\
 0 & 0 & 0 & \frac{(r-R_s) e^{-i\phi}}{2 r-R_s} \\
 0 & 0 & -\frac{(r-R_s) e^{i\phi}}{2 r-R_s} & 0
\end{array}
\right) ,
\end{eqnarray}
\begin{eqnarray}
\omega _3=\left(
\begin{array}{cccc}
 \frac{i C_{1}}{8 r-4 R_s} & \frac{R_s \sin 2\theta\; i e^{-i\phi}}{8 r-4 R_s} & 0 & 0 \\
 \frac{i R_s \sin 2\theta\; e^{i\phi}}{8r-4 R_s} & -\frac{i C_{1}}{8 r-4 R_s} & 0 & 0 \\
 0 & 0 & \frac{i C_{1}}{8 r-4 R_s} & \frac{R_s \sin 2\theta \;ie^{-i\phi}}{8 r-4 R_s} \\
 0 & 0 & \frac{i R_s  \sin 2\theta\; e^{i\phi}}{8 r-4 R_s} & -\frac{i C_{1}}{8 r-4 R_s}
\end{array}
\right) ,
\end{eqnarray}
where
\begin{eqnarray}
\noindent C_{1}&=& 4 r-3 R_s+R_s \cos(2 \theta ) .
\end{eqnarray}

\section{Gravitational Perturbation of The Relativistic Hydrogen Atom: The $1S_{1/2}$
States}\label{GRAVITATIONAL PERTURBATION}

From Eq. (\ref{DiracEq2})  the corresponding Hamiltonian in curved
space-time follows
\begin{eqnarray}
  H=-i \hbar c \Gamma_0 \Gamma^i  (\partial_i -\omega_{i}-iq A_{i})
  +i \hbar c (\omega_{0}+ i q A_0) -i mc^2 \Gamma_0 \label{H}.
\end{eqnarray}
The Dirac Hamiltonian in flat space is
\begin{eqnarray}
  H_0=-i\hbar c \gamma_0 \gamma^i  (\partial_i -iq A^{'}_i)
  - \hbar c  q A^{'}_0 -i mc^2 \gamma_0 \label{H0} ,
\end{eqnarray}
where $A^{'}_{\mu}$ are the electromagnetic vector potentials in
flat spacetime. Here we can take the approximation $A_{i}\cong
A^{'}_{i}=0$ and $A_{0}\cong A^{'}_{0}=-er^{-1}$, the detailed
discussions of this problem is contained in Ref.~\cite{Parker2}.
So the Hamiltonian of the gravitational perturbation is given by
\begin{eqnarray}
 H_I&=& H-H_0 \label{HI1}\nonumber\\
 &=&-i\hbar c \Gamma_0 \Gamma^i (\partial_i -\omega_{i})
  +i \hbar c \omega_{0}-i m_{e}c^2 \Gamma_0\nonumber\\
  & &+i\hbar  c\gamma_0 \gamma^i \partial_i
  +i m_{e}c^2 \gamma_0.
\end{eqnarray}

The exact solutions of the Dirac equation for a hydrogen atom in
flat space-time serve as the basis for perturbation theory. The
energy eigenvalues of a hydrogen atom are
\begin{eqnarray}
  E_{n\kappa}=m_{e}
  c^2\sqrt{1+\left(\frac{\zeta}{n-|\kappa|+s}\right)^2},
\end{eqnarray}
where $\zeta=Ze^2$, $s=\sqrt{\kappa^2-\zeta^2}$, $n=1,2,\cdots$ is
the principal quantum number.

The  bound state functions of a hydrogen atom can be written in
standard representation \cite{Rose,Strange} as
\begin{eqnarray}
  \psi=\psi^{M}_{\kappa}=\left( \begin{array}{c}g(r)\chi^{M}_{\kappa}\\
  -i f(r)\chi^{M}_{-\kappa}\end{array}\right),
\end{eqnarray}
here $M$ is the eigenvalue of $J_{z}$, $\kappa$ is the eigenvalue
of $K=\beta(\vec{\sigma}\cdot\vec{L}+I)$, the functions $f(r)$,
$g(r)$ and spinors $\chi^M_\kappa$, $\chi^M_{-\kappa} $ are given by
\begin{eqnarray}
f(r)&=&\frac{2^{s-\frac{1}{2}}\lambda^{s+\frac{3}{2}}}{\Gamma(2s+1)}\sqrt{\frac{\Gamma(2s+n_r+1)}{n_r!
\zeta K_c(\zeta K_c-\lambda\kappa )}}
\sqrt{1-\frac{W_{c}}{K_c}}r^{s-1}e^{-\lambda r} \nonumber\\
& &\left(\left(\kappa -\frac{\zeta K_c}{\lambda}
\right)\text{F}(-n_r, 2 s+1,2\lambda r)-n_r \text{F}(-n_r+1,2 s+1,
2\lambda r)\right),
\end{eqnarray}
\begin{eqnarray}
g(r)&=&-\frac{2^{s-\frac{1}{2}}\lambda^{s+\frac{3}{2}}}{\Gamma(2s+1)}\sqrt{\frac{\Gamma(2s+n_r+1)}{n_r!
\zeta K_c(\zeta K_c-\lambda\kappa )}}
\sqrt{1-\frac{W_{c}}{K_c}}r^{s-1} e^{-\lambda r} \nonumber\\
& &\left(\left(\kappa -\frac{\zeta K_c}{\lambda}
\right)\text{F}(-n_r, 2 s+1,2\lambda r)+n_r \text{F}(-n_r+1,2 s+1,
2\lambda r)\right),\\
\chi^M_\kappa& =&C_{1/2}Y^{M-1/2}_{l}\left(
  \begin{array}{c}
   1\\
   0
\end{array} \right)+C_{-1/2}Y^{M+1/2}_{l}\left(
  \begin{array}{c}
   0\\
   1
\end{array} \right),\\
 \chi^M_{-\kappa}& =&-C_{1/2}Y^{M-1/2}_{l}\left(
  \begin{array}{c}
   \cos\theta\\
   e^{i\phi}\sin\theta
\end{array} \right)-C_{-1/2}Y^{M+1/2}_{l}\left(
  \begin{array}{c}
   e^{-i\phi}\sin\theta \\
   -\cos\theta
\end{array} \right),
\end{eqnarray}
where $W_c=E_{n\kappa}/m_{e}c^2$, $K_c=m_{e}c^2/\hbar c$,
$\lambda=\sqrt{m_{e}^2c^4-E_{n\kappa}^2}/\hbar c$, $C_{1/2}$ and
$C_{-1/2}$ are the C-G coefficients.

For a hydrogen atom there are two
$1S_{1/2}(n=1,l=0,J=1/2,\kappa=-1)$ states, which  correspond
to $M=\pm 1/2$. The states can be written as

\begin{eqnarray}
\psi_{1}=\left(
\begin{array}{c}
 0 \\
 f(r) \\
 i g(r)  \sin\theta e^{ -i \phi }\\
 -i g(r) \cos\theta
\end{array} \right),
\end{eqnarray}
and
\begin{eqnarray}
\psi_{2}=\left(
\begin{array}{c}
 f(r)\\
 0 \\
 i g(r) \cos\theta
 \\
 i  g(r) \sin\theta  e^{i \phi}
\end{array}
\right),
\end{eqnarray}
where $\psi_{1}$ corresponds to $M=1/2$ and $\psi_{2}$ to
$M=-1/2$,
\begin{eqnarray}
f(r)&=&\frac{2^{-\frac{3}{2}+s} e^{-r \lambda } r^{-1+s} \lambda
^{\frac{1}{2}+s}\sqrt{K_{c}+W_{c}}  \sqrt{K_{c} \zeta
+\lambda }}{ K_{c}\sqrt{\pi \zeta   \Gamma(1+2 s)}},\\
g(r)&=&\frac{2^{-\frac{3}{2}+s} e^{-r \lambda } r^{-1+s} \lambda
^{\frac{1}{2}+s}\sqrt{K_{c}-W_{c}}  \sqrt{K_{c} \zeta +\lambda}}{ K_{c}\sqrt{\pi \zeta   \Gamma(1+2 s)}},
\end{eqnarray}
$\Gamma(1+2 s)$ is the $\Gamma$ function. The gravitational
perturbation matrix elements are
\begin{eqnarray}
\langle H_{I}\rangle_{ab}\equiv(\psi_{a},H_{I}\psi_{b}),
\end{eqnarray}
where the subscripts $a,b$ take on the values $1,2$. Because we take
the gravitational field metric as the Schwarzschild metric, so we
need to confirm the range of the integration. Here it is taken from
$R_n$ to $\infty$, $R_n\cong 1.3\times10^{-15}$ m is the atomic
nucleus radius. With the computer algebra system Mathematica, we
obtain the following results for those perturbation matrix elements
\begin{eqnarray}
\langle H_{I}\rangle_{ab}&=&-\frac{\delta_{ab}}{K_c^2 R_n \zeta  \Gamma(1+2 s)}2^{-1+2 s} c R_s \lambda  (R_n \lambda )^{2 s} (K_c \zeta +\lambda )\nonumber\\
& &\left( c m R_n W_c \text{E}_{1-2 s}(2 R_n \lambda)+
\sqrt{K_c^2-W_c^2}\hbar  \text{E}_{2-2 s}(2 R_n \lambda )\right),
\end{eqnarray}
where  $\text{E}_{n}(z)=\int^{\infty}_{1}e^{zt}/t^{n}dt$ is the
exponential integral function. Using the equation \cite{Parker2}
\begin{eqnarray}
  \text{det}[(\psi_a,H_{I}\psi_b)-E_i^{\;1}\delta_{ab}]=0\label{Ei},
\end{eqnarray}
from the usual perturbation theory of a degenerate energy
eigenvalue,  it follows that both of the degenerate $1S_{1/2}$
levels are shifted by the same perturbation:
\begin{eqnarray}
E^{1}(1S_{1/2})&=&-\frac{1}{K_c^2 R_n \zeta  \Gamma(1+2 s)}2^{-1+2 s} c R_s \lambda  (R_n \lambda )^{2 s} (K_c \zeta +\lambda )\nonumber\\
& &\left( c m R_n W_c \text{E}_{1-2 s}(2 R_n \lambda)+
\sqrt{K_c^2-W_c^2}\hbar  \text{E}_{2-2 s}(2 R_n \lambda )\right)
.\label{E1}
\end{eqnarray}
Substituting the constant values in Table 1 into Eq. (\ref{E1}),
we get
\begin{eqnarray}
  E^{1}(1S_{1/2})=-1.19956\times10^{-38} \; \text{ev}.
\end{eqnarray}

\begin{tabular}{llll}
\multicolumn{4}{c}{TABLE 1.  The constants table \cite{Eidelman}  } \\
\hline\hline
 Quantity  & Symbol  & Value &Units \\
electron charge magnitude &  $e $ & 1.60217653 $\times 10^{-19}$& C\\
speed of light in vacuum & $c $ & 2.99792458 $\times10^{-8}$&m\; s$^{-1}$\\
electron mass &$m_{e}$ & 9.91093826$\times 10^{-31}$&kg \\
Planck constant, reduced &$\hbar$ & 1.05457168$\times 10^{-34}$&J\;s \\
permittivity of free space &$\epsilon_{0}$ &8.854187817
$\times10^{-12}$ & s$^{4}$\;A$^{2}$\;kg$^{-1}$\;m$^{-3}$\\
proton mass &$ M_{p} $& 1.67262171 $\times10^{-27} $&  kg\\
gravitation constant & $G $ &6.6742$\times10^{-11}$ & m$^3$
kg$^{-1}$s$^{-2}$\\
\botrule
\end{tabular}

\section{Summary and Discussion}\label{SUMMARY}
In a similar calculation as the $1S_{1/2}$ state, we find that all
the relativistic $1S$, $2S$, $2P$, $3S$, $3P$, $3D$, $4S$ and $4P$
energy levels are respectively shifted as the same amount listed in
Table 2. This means that the first order gravitational perturbations
can partly remove the degeneracy of the hydrogen atom states.
Although the effect is very small, but form Table 2 we find that the
quantity of corrections of the energy levels with same principal
quantum number $n$ and total angular momentum quantum number $J$,
like $2S_{1/2}$ and $2P_{1/2}$, $3S_{1/2}$ and $3P_{1/2}$,
$3P_{3/2}$ and $3D_{3/2}$, are very closely. But for the levels with
same principal quantum number
 and different total angular momentum quantum number, like
$3S_{1/2}$ and $3P_{3/2}$, their corrections have obvious
difference. Those calculations show that the gravitational
corrections are sensitive to the total angular momentum quantum
number. It is a very important feature of the interaction between
gravitational fields and spinor fields.  With this feature we can
find the gravitational effect in other system, and make a test of
general relativity at the quantum level.

\begin{tabular}
{ll}
\multicolumn{2}{c}{TABLE 2.  The energy-level shifts  } \\
\hline\hline
State  & The energy-level shift (Unit: ev)  \\
\hline
$1S_{1/2}$&  -1.19956 $\times10^{-38}$\\
$2S_{1/2}$&  -8.99637 $\times10^{-39}$\\
$2P_{1/2}$&  -8.99562 $\times10^{-39}$\\
$2P_{3/2}$&  -2.99862 $\times10^{-39}$\\
$3S_{1/2}$&  -6.66389 $\times10^{-39}$\\
$3P_{1/2}$&  -6.66353 $\times10^{-39}$\\
$3P_{3/2}$&  -2.66544 $\times10^{-39}$\\
$3D_{3/2}$&  -2.66538 $\times10^{-39}$\\
$4S_{1/2}$&  -5.24777 $\times10^{-39}$\\
$4P_{1/2}$&  -5.24756 $\times10^{-39}$\\
\botrule
\end{tabular}


\begin{thebibliography}{99}


\bibitem{Parker}
L. Parker, Phys. Rev. Lett. \textbf{44} (1980) 1559.

\bibitem{Parker2}
L. Parker, Phys. Rev. D\textbf{22} (1980) 1922.

\bibitem{Parker3}
L. Parker and  L. O. Pimentel, Phys. Rev. D\textbf{25} (1982)
3180.

\bibitem{Fischbach}
 Y. S. Duan,
 J. Exptl. Theoret. Phys. (U.S.S.R.) \textbf{34} (1958) 632;
 E. Fischbach and B. S. Freeman,
 Phys. Rev. D\textbf{23} (1981) 2157.

\bibitem{Marques}
G. de A. Marques and V. B. Bezerra, Phys. Rev. D\textbf{66} (2002)
105011.

\bibitem{Fischer}
M. Fischer et al., Phys. Rev. Lett. \textbf{92} (2004) 230802.

\bibitem{Kolachevsky}
N. Kolachevsky, J. Alnis, S. D. Bergeson, and T. W. H\"{a}nsch,
Phys. Rev. A\textbf{73} (2006) 021801.



\bibitem{Cesar}
C. L. Cesar et al., Phys. Rev. Lett. \textbf{77} (1996) 255.

\bibitem{Killian}
T. C. Killian et al., Phys. Rev. Lett. \textbf{81} (1998) 3807.


\bibitem{Jentschura}
U. D. Jentschura, P. J. Mohr, and G. Soff, Phys. Rev. Lett.
\textbf{82} (1998) 53.


\bibitem{Kibble}
T. W. B. Kibble, J. Math. Phys. \textbf{2} (1961) 212.

\bibitem{Loll}
R. Loll, \emph{Discrete Approaches to Quantum Gravity in Four
Dimensions},\\ (http://www.livingreviews.org/lrr-1998-13).

\bibitem{Poisson}
E. Poisson, \emph{The Motion of Point Particles in Curved
Spacetime},\\
(http://www.livingreviews.org/lrr-2004-6).

\bibitem{Misner}
C. W. Misner, K. S. Thorne, and J. A. Wheeler, \emph{Gravitation},
Freeman, San Francisco (1973), p.332.

\bibitem{Schwinger}
J. Schwinger, Phys. Rev. \textbf{130} (1963) 1253.

\bibitem{Maluf}
J. W. Maluf, J. F. da Rocha-Neto, T. M. L. Toribio, and K. H.
Castello-Branco, Phy. Rev. D\textbf{65} (2002) 124001.


\bibitem{Rose}
M. E. Rose, \textit{Relativistic Electron Theory}, Wiley, New York
(1961), p.177.

\bibitem{Strange}
P. Strange, \textit{Relativistic Quantum Mechanics}, Cambridge
University Press, Cambridge (1998), p.229.








\bibitem{Eidelman}
S. Eidelman, et al., Phys.
Lett. B\textbf{1} (2004) 592.


\end{thebibliography}
\end{document}